\documentclass[12pt]{article} % Generic single-column article class
\usepackage[margin=1in]{geometry} % Adjust margins
\usepackage{graphicx}   % For including images
\usepackage{amsmath}    % For mathematical symbols
\usepackage{amssymb}    % For additional symbols
\usepackage{hyperref}   % For clickable references
\graphicspath{{figures2/}}
\usepackage{caption}    % Improved captions
\usepackage{subcaption} % Subfigures
\usepackage{float}      % Control figure placement
\usepackage{setspace}   % Line spacing
\usepackage{enumitem}   % Better control over lists
\usepackage{algorithm}
\usepackage{algpseudocode}
\usepackage{graphicx}
\usepackage{array}
\usepackage{caption}
\usepackage[table]{xcolor}
\usepackage{amsmath}
\usepackage{booktabs}
\usepackage{graphicx}
\usepackage{array}
\usepackage{fancyhdr} % For custom headers and footers

\begin{document}
	
	\title{\textbf{PIXHELL: When Pixels Learn to Scream}}
	\author{
		Mordechai Guri \\
		Ben-Gurion University of the Negev, Israel \\
		Department of Software and Information Systems Engineering \\
		Offensive-Cyber Research Lab  \\
		\texttt{gurim@post.bgu.ac.il} \\ 
		\url{https://www.covertchannels.com} \\
		Demo video: \url{https://www.instagram.com/reel/C\_0Osn6JwHJ/}
	}
	\date{\today}
	
	\maketitle
	
	\begin{abstract}
	This paper presents a technique for generating sound by leveraging the electrical properties of liquid crystal displays (LCDs). The phenomenon occurs due to vibrational noise produced by capacitors within the LCD panel during rapid pixel state transitions. By modulating these transitions through specially crafted bitmap patterns projected onto the screen, we demonstrate how weak yet audible acoustic signals can be generated directly from the display. We designed, implemented, evaluated, and tested a system that repurposes the LCD as a sound-emitting device. Potential applications for this technique include low-power auditory feedback systems, short-range device communication, air-gap covert channels, secure auditory signaling, and innovative approaches to human-computer interaction.

	\end{abstract}
	
	\section{Introduction}	
	Audio hardware is integral to modern computing, supporting functions like multimedia playback, system alerts, and assistive tools. Today’s computers typically include dedicated audio components such as sound cards, digital-to-analog converters (DACs), and speakers. These elements are seamlessly integrated into hardware and software systems, creating a reliable framework for generating and transmitting high-quality audio.
	
	In practice, not all computers are equipped with functional audio hardware or accessible speakers. For example, industrial PCs, embedded devices, and ultra-thin laptops often lack built-in speakers due to cost or design constraints. In secure environments, such as air-gapped systems, audio hardware may also be disabled or removed to prevent covert data leaks, a measure commonly referred to as 'audio-gap' security \cite{AirGapCo49:online}. Additionally, even when audio hardware is present, its use can be restricted by operating system (OS) policies, missing drivers, or permissions that block access to audio APIs. These issues highlight significant challenges in generating audio output, particularly in constrained or secure contexts.
	
	In this paper, we present a method for generating sound by repurposing liquid crystal displays (LCDs) as sound-emitting devices. This method takes advantage of the vibrational noise produced by capacitors and other electronic components during rapid pixel state transitions, enabling the LCD to function as a basic speaker capable of emitting audio signals. By displaying specially generated pixel patterns, or bitmaps, on the screen and designing them based on the specific characteristics of the display, such as resolution and refresh rates, we demonstrate how these transitions can produce sound. Furthermore, we show that by modulating these pixel transitions, information can be encoded and transmitted through the generated audio. This approach also offers an alternative mechanism for audio generation that bypasses traditional audio hardware, providing a practical solution for scenarios where conventional methods are unavailable or restricted.
	
	Notably, in our prior academic work, PIXHELL \cite{guri2024pixhell}, we explored the use of LCDs to generate sound specifically in the context of air-gap security. The current work broadens the focus, presenting a comprehensive technical exploration that highlights the underlying mechanisms and examines a wide range of applications for LCD-based audio communication beyond security concerns.
			
	\begin{figure}[htbp]
		\centering
		\includegraphics[width=\linewidth]{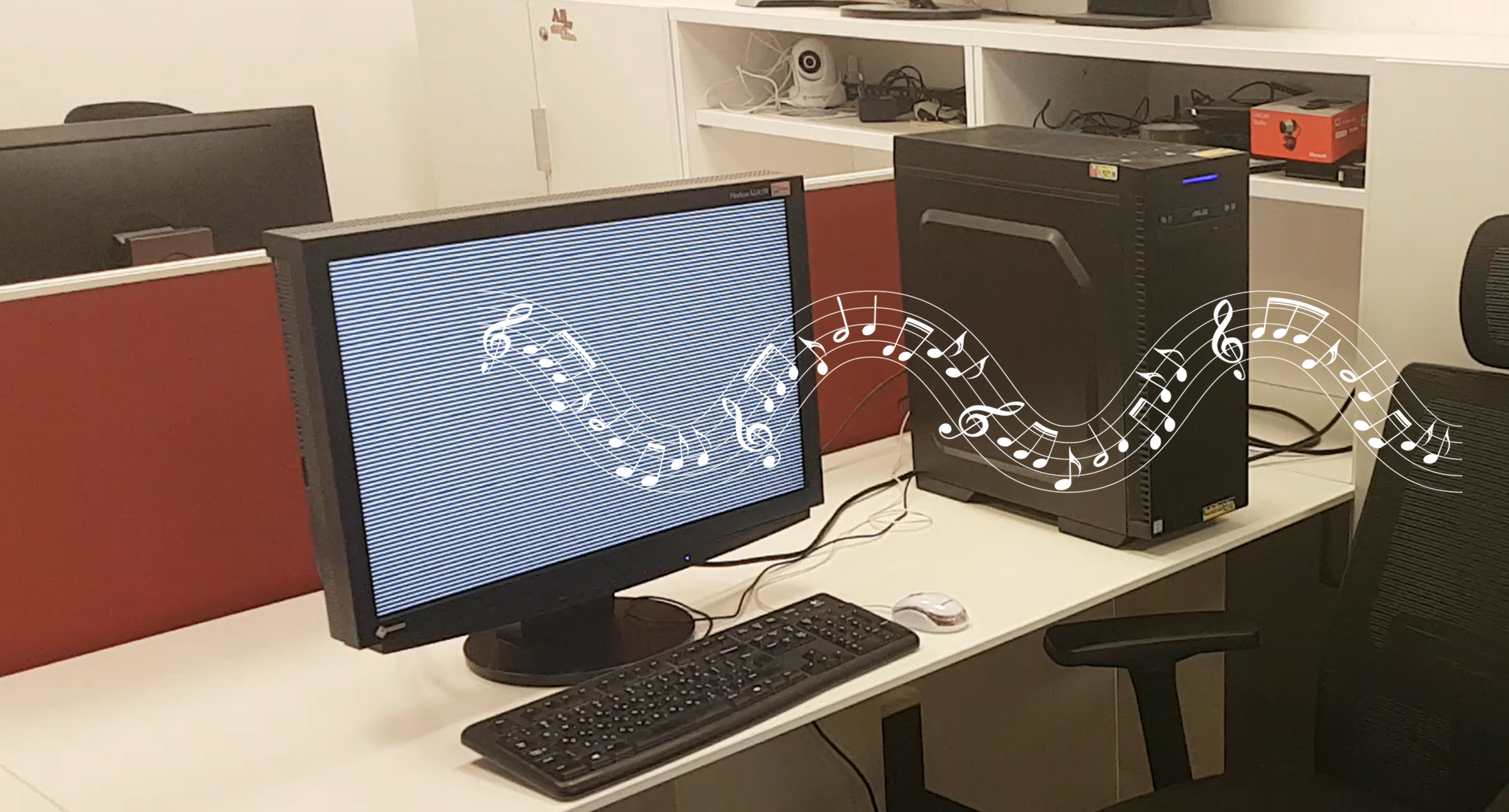}
		\caption{The `Screaming Pixels' Phenomenon. An illustration of the acoustic waves generated by the visual pixel patterns displayed on the LCD screen.}
		\label{fig:lcdsound}
	\end{figure}
	 Figure \ref{fig:lcdsound} illustrates the concept of `screaming pixels,' acoustic waves generated by the visual pixel patterns displayed on the LCD screen.

	\subsection{Use Cases for LCD-Based Audio Communication}
	This method offers several applications in legitimate contexts, leveraging the ability of screens to generate sound for communication. Despite the weak and short-range nature of the acoustic signal, it can enable short-range communication systems where screens act as transmitters and microphones or other sound sensors serve as receivers. This makes it particularly useful in controlled environments with specific communication requirements. 
	The approach could also serve as a low-cost solution for device pairing, authentication, or data sharing in consumer electronics. Acoustic signals could be used to securely authenticate devices in proximity, eliminating the need for complex hardware or wireless protocols. Additionally, it could inspire innovative applications in interactive displays, educational tools, and assistive technologies, where sound-based feedback is desired without relying on conventional audio hardware.
	The ability to produce controlled or inaudible sounds from a screen may also present opportunities for misuse. For instance, it could be exploited to create a covert channel for data exfiltration, transmitting information acoustically to an external receiver, even in environments lacking audio hardware.
		
\section{Related Work}
Sound has long been used as a communication medium, with conventional applications in technologies such as line modems and fax machines, which rely on acoustic signals for data transmission \cite{hanes2008fax}. These systems encode information into audible or inaudible frequencies, demonstrating the foundational principles of sound-based communication. In more modern contexts, acoustic communication has expanded to include techniques leveraging the omnipresence of microphones and speakers in everyday environments, facilitating innovative applications like device pairing and covert data channels \cite{hanspach2014recent}. 

Audio-based communication has been extensively studied in computer science, particularly in the contexts of communication channels, covert channels, data exfiltration, and human-computer interaction. Techniques utilizing sound as a communication medium often exploit the inherent advantages of acoustic signals, such as their ability to traverse through the air and their compatibility with wide range of existing hardware. For instance, ultrasonic communication has been explored for device pairing (e.g., smartphones), leveraging inaudible frequencies to enable short-range, high-frequency data exchanges \cite{getreuer2017ultrasonic}. Other approaches have demonstrated the feasibility of both audible and inaudible audio channels for covert communication channels \cite{wong2018crossing}. Note that covert channels, in general, have long been recognized as a significant threat due to their ability to circumvent traditional security measures. Recent research has explored their exploitation in Internet protocols, highlighting their persistent impact on security systems \cite{Caviglione2024, Zorawski2023}.

\subsection{Active Acoustic Side-Channels}
Researchers have explored innovative ways to generate sound from systems that lack dedicated audio hardware. Guri et al presented Fansmitter, a malware that manipulates CPU and GPU fan speeds to produce acoustic signals detectable by nearby devices such as smartphones \cite{guri2020fansmitter}\cite{guri2022gpu}. Similarly, DiskFiltration encodes data into the operational noise of hard disk drives (HDDs) actuator arm during read and write operations, allowing for acoustic data transmission \cite{guri2016diskfiltration}. Another method, CD-LEAK, leverages the mechanical noise of optical drives, such as CD/DVD players, for data modulation \cite{guri2020cd}. Additionally, power-supply-based attacks utilize the switching frequencies of power supply units (PSUs) to emit sound or ultrasonic waves, providing a novel channel for data exfiltration without relying on conventional audio hardware \cite{guri2021power}. More recently, EL-GRILLO demonstrated how ultrasonic data can be leaked from air-gapped PCs using the tiny motherboard buzzer, further expanding the field of acoustic side-channel research \cite{guri2023grillo}. 
Unconventional sound sources have further expanded the field of audio-based communication. For instance, InkFiltration demonstrates the use of mechanical noise from inkjet printers to transmit data over short distances \cite{de2022inkfiltration}. The MOSQUITO method employs ultrasonic communication to transfer data between air-gapped systems, enabling speaker-to-speaker communication using inaudible frequencies \cite{guri2018mosquito}. More recently, PIXHELL has explored the potential of generating sound through LCD displays by modulating pixel transitions \cite{guri2024pixhell}. While PIXHELL introduces the concept of LCD-based audio generation, it primarily emphasizes the broader possibilities rather than delving into the precise algorithms and methods required for effective sound production and data transmission.

\subsection{Passive Acoustic Side-Channels}
Passive Side-channel attacks exploit unintended information leakage from devices, such as timing, power consumption, or acoustic emanations. Among these, acoustic side-channels focus on analyzing sound emitted by devices to infer operations or extract sensitive information. These attacks are often passive, requiring no direct interaction with the device, and are particularly effective when other side-channels are inaccessible. Backes et al. demonstrated an attack leveraging the acoustic emanations of dot matrix printers to reconstruct printed text, highlighting vulnerabilities in printer operations \cite{Backes2010}. Deepa et al. provided a comprehensive survey on acoustic side-channel attacks, detailing various techniques and methodologies to exploit sound leakage for data retrieval \cite{Deepa2013}. Faruque et al. introduced an acoustic side-channel attack on 3D printing systems, where sound emanations were used to infer the geometry of printed objects, bridging the physical and cyber domains \cite{Faruque2016}. Halevi and Saxena analyzed keyboard acoustic emanations to eavesdrop on password entry, employing statistical and machine learning models to reconstruct typed input \cite{Halevi2015}. Taheritajar et al. reviewed advancements in keyboard acoustic side-channel attacks, emphasizing improvements in sound recording techniques and their implications for security \cite{Taheritajar2023}. Harrison et al. proposed a deep learning-based method for acoustic side-channel attacks on keyboards, achieving significant improvements in accuracy over traditional approaches \cite{Harrison2023}. Toreini et al. explored the feasibility of acoustic side-channel attacks on the Enigma machine, showing that its operations could be inferred through sound leakage \cite{Toreini2015}. Finally, Shumailov et al. introduced an acoustic side-channel attack targeting virtual keyboards on smartphones, reconstructing typed input by analyzing the sound of touchscreen taps \cite{Shumailov2019}.

\section{Technical Background}
\label{sec:1}
Liquid crystal displays (LCDs) are one of the most widely used display technologies, found in everything from consumer electronics to industrial equipment. LCDs operate by manipulating liquid crystals to control the passage of light, creating visible images. This section provides an overview of LCD screen types and their internal components, followed by a discussion of potential noise sources within these displays.

\subsection{LCD Screen Types}
LCD displays come in a variety of types, primarily differentiated by their backlighting and pixel addressing mechanisms. Twisted Nematic (TN) panels are known for their fast response times and cost efficiency, though they are often limited by poor color accuracy and narrow viewing angles \cite{kim2005unique}. In contrast, In-Plane Switching (IPS) panels provide superior color reproduction and wide viewing angles, making them a popular choice for professional monitors and high-end devices \cite{hong2007plane}. Vertical Alignment (VA) panels are characterized by high contrast ratios, offering deep blacks and vibrant colors, which are particularly suitable for applications requiring enhanced image quality \cite{park2010pixel}. Another distinction lies in the addressing mechanisms: passive matrix LCDs utilize a basic conductive grid, while active matrix displays, such as thin-film transistor LCDs (TFT-LCDs), incorporate transistors at each pixel, resulting in better control and image sharpness. Among these, IPS and VA panels dominate modern computer monitors and televisions due to their balance of image quality, color fidelity, and viewing angles, making them the most widely adopted technologies in current consumer electronics.

\subsection{Internal Components}
Internally, an LCD screen comprises several components:

\begin{itemize}
	\item \textbf{Backlight Unit (BLU).} Provides uniform illumination for the display, typically achieved using an array of light-emitting diodes (LEDs) \cite{kobayashi2009lcd}.
	\item \textbf{Liquid Crystal Layer.} Modulates light by altering the alignment of liquid crystal molecules in response to applied electric fields.
	\item \textbf{Polarizers.} Linear polarizing filters that control light orientation to enhance contrast and produce visible images.
	\item \textbf{Driver Circuitry.} Includes thin-film transistors (TFTs) and integrated circuits responsible for controlling pixel-level operations by precisely adjusting the voltage across each liquid crystal cell.
	\item \textbf{Capacitors and Power Supply.} Ensure stable voltage levels for maintaining consistent pixel states and powering the display’s active components.
\end{itemize}

The process of image rendering on an LCD screen begins with the transmission of a video signal from the source device (e.g., a computer or media player), via digital interfaces like HDMI or DisplayPort. The signal contains encoded pixel data and synchronization information. This data is processed by the timing controller (TCON), which converts the digital signal into appropriate drive signals for the gate and source drivers. The gate drivers control the rows of pixels, activating the appropriate liquid crystal cells, while the source drivers apply the necessary voltages to align the liquid crystal molecules. The liquid crystal layer modulates light intensity for each pixel, which is illuminated by the backlight. Polarizers filter and align the light, ensuring accurate color reproduction. Synchronization signals, such as horizontal sync (h-sync) and vertical sync (v-sync), coordinate the pixel data for each line and frame to ensure proper rendering.

The refresh rate, measured in hertz (Hz), determines how often the screen updates per second, influencing the smoothness of motion, especially in fast-moving content like video games or action sequences. Modern LCD displays offer a range of refresh rates, from the standard 60Hz to higher options like 120Hz and 240Hz, which provide a significantly smoother visual experience.

In addition to refresh rate, resolution has seen substantial advancements, moving from standard definition (SD) to high definition (HD), full HD (FHD, 1920x1080), 4K Ultra HD (3840x2160), and even 8K (7680x4320). The increase in pixel density enhances image clarity and detail, especially on large screens, where lower resolutions might otherwise lead to noticeable pixelation.

\subsection{Noise Sources in LCD Screens}
LCD screens, although designed primarily for visual output, can unintentionally generate acoustic noise due to the behavior of their internal components. The primary source of this noise is capacitor squeal, which occurs in the power supply unit. This phenomenon is driven by the piezoelectric effect, where capacitors vibrate under alternating electric fields. These vibrations, especially during rapid changes in control voltage caused by pixel transitions, result in mechanical collisions with the printed circuit board (PCB), producing audible sounds \cite{TDK2006}.

The electrostriction effect further amplifies this behavior. Capacitors compress or expand in response to electric field variations, generating additional mechanical vibrations. Voltage fluctuations, particularly during pixel transitions such as switching between black and white states, intensify these vibrations, leading to stronger acoustic emissions. These transitions are reflected in the varying voltage demands of the power supply, further increasing noise levels \cite{TI2016,Kim2019}.

Another significant contributor is the switching power supply, which regulates voltage for various components, including the backlight and driver circuitry. The rapid toggling of electrical currents in these power supplies produces high-frequency oscillations that can induce vibrations in nearby elements, contributing to the overall noise profile. 

Additional noise may also arise from the driver circuitry, which translates pixel-level RGB values into electrical signals. This process continuously adjusts control voltages that influence capacitor behavior. Furthermore, elements such as thin-film transistors (TFTs), the backlight driver circuits, and even the liquid crystal layer itself during rapid alignment changes can contribute to the acoustic emissions, though their impact is typically less pronounced \cite{MDPI2022}.

The combined effects of capacitor squeal, electrostriction, voltage variations, and switching power supplies make LCD screens a potential source of acoustic noise. While these emissions are usually negligible, they can be exploited for sound generation by deliberately modulating the display’s control signals. This understanding forms the basis for the sound generation techniques explored in this work.

\section{Signal Generation}
There exists a direct correlation between the acoustic noise emitted by an LCD screen and the visual content displayed on it. This relationship is primarily governed by the electrical activity and voltage changes in the screen's components during image rendering. When an image is presented on an LCD, the driver circuitry modulates the control voltages applied to the liquid crystal pixels. These voltage changes, in turn, influence the capacitors and power supply, causing them to emit vibrations at frequencies tied to the pixel activity. 

By carefully crafting the displayed image, it is possible to control the specific frequencies of the noise generated by the screen. For example, switching a large region of pixels between states (e.g., from black to white or to intermediate grayscale values) imposes a significant and rapid demand on the power supply, creating predictable oscillations. The spatial and temporal characteristics of the image, such as patterns, gradients, or rapid changes in intensity, can be fine-tuned to induce desired acoustic frequencies.

This ability to modulate noise stems from the fact that the power consumption and voltage adjustments required to display an image vary with its complexity and content. High-contrast transitions, alternating patterns, and dynamic changes in pixel intensity result in distinct power fluctuations that are reflected in the emitted noise. By designing an image sequence with specific visual patterns, the frequency spectrum of the emitted noise can be shaped, effectively turning the LCD into a controllable sound-emitting device.

In this paper, we leverage this phenomenon by developing algorithms that generate carefully designed images to produce targeted frequencies. This approach allows for precise modulation of the acoustic emissions, enabling the transmission of audio signals or encoded data through the LCD screen. The following sections describe the methodology and implementation of the image-crafting algorithms that facilitate this control.

\subsection{The Pixel Clock}
To understand our approach to generating bitmap patterns for controlling acoustic emissions from an LCD, we first define a few key terms and calculations that are central to the algorithm.

\textit{Resolution} denotes the total number of pixels in an LCD screen, expressed as $\text{Width} \times \text{Height}$. For example, a $1920 \times 1080$ resolution means 1920 pixels horizontally and 1080 pixels vertically. The total number of pixels is given by:
\[
\text{Total Pixels} = \text{Width} \times \text{Height}.
\]

\textit{Refresh rate}, measured in Hertz (Hz), indicates how frequently the screen updates its displayed content per second. For instance, a refresh rate of 60 Hz means the screen refreshes 60 times per second, which is critical for ensuring smooth motion in dynamic content.

\textit{Pixel clock} denotes the rate at which pixel data is transmitted to the screen. This parameter, typically measured in kilohertz (kHz) or megahertz (MHz), is influenced by the total number of pixels updated per second, including blanking intervals used for synchronization purposes. The pixel clock is computed as follows:
\[
\text{Pixel Clock (Hz)} = \text{Total Pixels} \times \text{Refresh Rate}.
\]

\textit{Blanking intervals} refer to non-visible periods during each frame update when no pixel data is displayed. These intervals are required to allow the display hardware to reset and prepare for the next line or frame of pixels. They include horizontal blanking, which occurs at the end of each line, and vertical blanking, which occurs at the end of each frame. Although blanking intervals do not contribute directly to the visible content, they introduce overhead in terms of pixel transmission. The exact percentage of overhead, denoted as $\beta$, depends on the specific display interface and can vary; in many cases, it is approximately 10\%. Therefore, the pixel clock with blanking intervals included is given by:
\[
\text{Pixel Clock (Hz)} = \text{Width} \times \text{Height} \times \text{Refresh Rate} \times (1 + \beta),
\]
where $\beta$ represents the blanking overhead ratio (e.g., $\beta = 0.1$ for 10\% overhead).

\subsection{PWM and Square Waves}
Pulse Width Modulation (PWM) is a fundamental technique used in digital signal processing and hardware control to encode information or generate signals. In the context of square wave generation, PWM allows the modulation of a digital signal's duty cycle to achieve desired properties such as frequency and amplitude. The generated square wave can then drive transitions that affect the emitted noise from the LCD.

%\paragraph{Square Wave Definition} 
A square wave is a periodic waveform that alternates between two levels, typically high (\(V_{\text{high}}\)) and low (\(V_{\text{low}}\)). Mathematically, a square wave \(S(t)\) of frequency \(f\) and duty cycle \(D\) is defined as:
\[
S(t) = 
\begin{cases} 
	V_{\text{high}}, & 0 \leq (t \mod T) < D \cdot T \\
	V_{\text{low}}, & D \cdot T \leq (t \mod T) < T
\end{cases}
\]
where \(T = \frac{1}{f}\) is the period of the wave, \(D \in [0, 1]\) is the duty cycle, and \(t\) is time.

%\paragraph{PWM and Duty Cycle} 
PWM encodes information by adjusting the duty cycle \(D\) of a digital signal. The duty cycle represents the fraction of time the signal is high within one period:
\[
D = \frac{t_{\text{high}}}{T},
\]
where \(t_{\text{high}}\) is the duration the signal remains high during a single period \(T\). For a 50\% duty cycle (\(D = 0.5\)), the square wave spends an equal amount of time in the high and low states, producing a balanced waveform.

\paragraph{Frequency and Pixel Clock Correlation}
In our context, the square wave's frequency is directly linked to the pixel clock (\(f_{\text{pixel}}\)) and the image pattern displayed on the screen. The wave frequency \(f_{\text{wave}}\) is determined by:
\[
f_{\text{wave}} = \frac{f_{\text{pixel}}}{\text{cycleSize}},
\]
where \(\text{cycleSize}\) is the number of pixels corresponding to one complete wave cycle:
\[
\text{cycleSize} = \frac{f_{\text{pixel}}}{f_{\text{wave}}}.
\]
For a square wave with a frequency of \(f_{\text{wave}}\), half of the pixels in one cycle (\(\frac{\text{cycleSize}}{2}\)) represent the high state, and the other half represent the low state.

\paragraph{Impact on LCD Noise}
The rapid pixel transitions generated by the square wave modulate the electrical signals within the LCD. These transitions induce vibrations in capacitors and the internal power supply, producing acoustic emissions at the wave frequency. By carefully choosing \(f_{\text{wave}}\) and designing the bitmap pattern accordingly, it is possible to control the frequency and characteristics of the emitted noise.

This mathematical foundation establishes the principles behind the bitmap generation algorithm described in the following subsection.

\subsection{Bitmap Generation}
The relationship between the pixel clock and the acoustic noise emitted by an LCD screen demonstrates how pixel update rates influence noise frequency. This frequency, dictated by the pixel clock, can be controlled through the careful design of the displayed image. By crafting a bitmap with a square wave pattern, it is possible to create predictable pixel transitions that correspond to specific acoustic frequencies.

We developed an algorithm that generates such a bitmap based on the pixel clock and a target carrier frequency. When this pattern is displayed, it leverages the inherent noise generation properties of the LCD screen to emit acoustic signals at the desired frequency. This work highlights how visual content can be systematically designed to modulate acoustic output.

\begin{algorithm}[H]
	\caption{\textbf{Generate Square Wave Bitmap}}
	\label{alg:generate_bitmap}
	\begin{algorithmic}[1]
		\Require Pixel clock (\texttt{pixelClock}) in kHz, carrier frequency (\texttt{freq}) in Hz, image dimensions (\texttt{width}, \texttt{height})
		\Ensure A 2D bitmap (\texttt{bitmap}) representing a square wave pattern
		
		\State \textbf{Calculate cycle size and half-cycle:}
		\Statex \hspace{\algorithmicindent} $pixelClockHz \gets pixelClock \times 1000$
		\Statex \hspace{\algorithmicindent} $cycleSize \gets \frac{pixelClockHz}{freq}$
		\Statex \hspace{\algorithmicindent} $halfCycle \gets \frac{cycleSize}{2}$
		
		\State \textbf{Initialize bitmap:}
		\Statex \hspace{\algorithmicindent} \texttt{bitmap} $\gets$ Create 2D array of size \texttt{height} $\times$ \texttt{width}
		
		\State \textbf{Generate pixel values:}
		\State $sampleNumber \gets 0$
		\For{$y \gets 0$ \textbf{to} \texttt{height} $- 1$}
		\For{$x \gets 0$ \textbf{to} \texttt{width} $- 1$}
		\State $remainder \gets sampleNumber \mod cycleSize$
		\If{$remainder < halfCycle$}
		\State \texttt{bitmap}[$y$][$x$] $\gets 255$ 
		\Else
		\State \texttt{bitmap}[$y$][$x$] $\gets 0$ 
		\EndIf
		\State $sampleNumber \gets sampleNumber + 1$
		\EndFor
		\EndFor
		
		\State \Return \texttt{bitmap}
	\end{algorithmic}
\end{algorithm}

Algorithm~\ref{alg:generate_bitmap} generates a bitmap that encodes a square wave pattern based on the pixel clock and desired carrier frequency. The process involves the following steps:

\begin{enumerate}
	\item \textbf{Calculate Cycle Size:} The cycle size determines how many pixels correspond to one complete wave period, based on the relationship:
	\[
	\text{cycleSize} = \frac{\text{pixelClock (Hz)}}{\text{freq (Hz)}}.
	\]
	The \texttt{halfCycle} represents the number of pixels in the high or low state of the wave.
	
	\item \textbf{Initialize Bitmap:} A 2D array of the specified dimensions (\texttt{width} $\times$ \texttt{height}) is created to store the grayscale pixel values.
	
	\item \textbf{Generate Pixel Values:} The algorithm iterates over each pixel in the 2D array. The current pixel's position in the wave cycle (\texttt{remainder}) is calculated using the modulo operation. If the remainder is within the \texttt{halfCycle}, the pixel is set to 255 (white); otherwise, it is set to 0 (black). The \texttt{sampleNumber} variable ensures that the wave pattern continues seamlessly across rows.
	
	\item \textbf{Output Bitmap:} The generated bitmap encodes a square wave pattern, where the frequency of transitions between black and white corresponds to the carrier frequency.
\end{enumerate}

The resulting bitmap can be displayed on an LCD screen to produce acoustic emissions at the specified frequency, leveraging the inherent noise generation properties of the display.

\begin{figure}[H]
	\centering
	\includegraphics[width=0.8\linewidth]{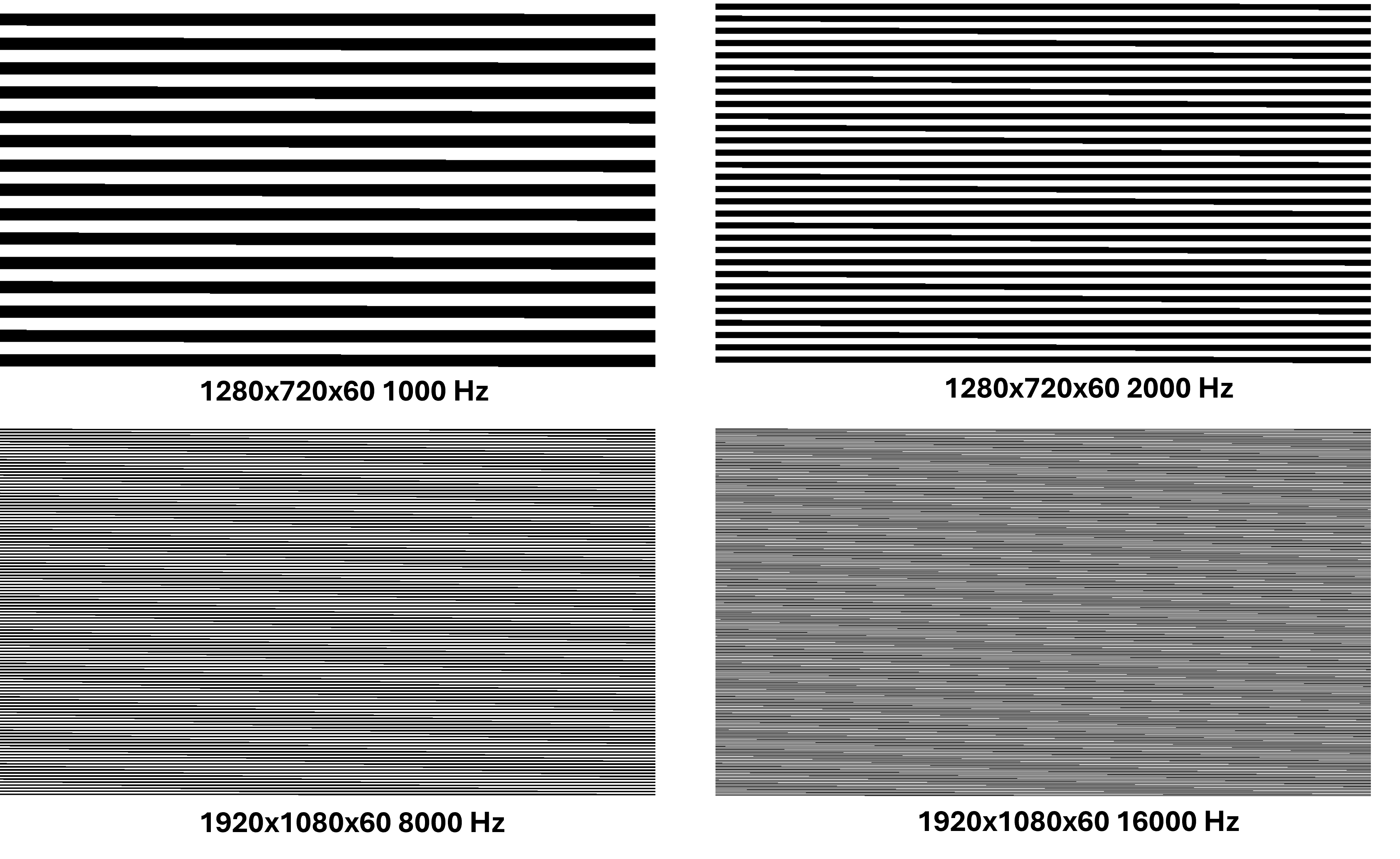}
	\caption{Four bitmaps generated by the algorithm with different parameters; resolution, pixel clock, and frequency. Clockwise direction: 1280$\times$720@60Hz 1000 Hz, 1280$\times$720@60Hz 2000 Hz, 1920$\times$1080@60Hz 8000 Hz, and 1920$\times$1080@60Hz 16000 Hz.}
	\label{fig:bitmap_examples}
\end{figure}

\paragraph{Duty Cycle Assumption} 
In this algorithm, the duty cycle \(D\) is fixed at 50\%. This means the square wave alternates equally between high (\(V_{\text{high}}\)) and low (\(V_{\text{low}}\)) states within each cycle, creating a balanced waveform. Mathematically, this corresponds to:
\[
D = \frac{t_{\text{high}}}{T} = 0.5.
\]
The high and low states each occupy half of the cycle:
\[
t_{\text{high}} = t_{\text{low}} = \frac{T}{2}.
\]
This simplification ensures predictable and consistent frequency generation while maintaining efficient use of the pixel transitions for sound modulation.

\noindent Algorithm~\ref{alg:generate_bitmap_duty_cycle} extends the previous algorithm by introducing a configurable duty cycle parameter \(D\). While the earlier implementation assumed a 50\% duty cycle (equal high and low states), this version allows the duty cycle to be adjusted, enabling asymmetric square waves. The duty cycle \(D\) determines the proportion of the cycle spent in the high state, where \(0 < D < 1\).

\begin{algorithm}[H]
	\caption{\textbf{Generate Square Wave Bitmap with Configurable Duty Cycle}}
	\label{alg:generate_bitmap_duty_cycle}
	\begin{algorithmic}[1]
		\Require Pixel clock (\texttt{pixelClock}) in kHz, carrier frequency (\texttt{freq}) in Hz, duty cycle (\texttt{D}), image dimensions (\texttt{width}, \texttt{height})
		\Ensure A 2D bitmap (\texttt{bitmap}) representing a square wave pattern with duty cycle \texttt{D}
		\State \textbf{Calculate cycle size and high/low pixels:}
		\Statex \hspace{\algorithmicindent} $pixelClockHz \gets pixelClock \times 1000$
		\Statex \hspace{\algorithmicindent} $cycleSize \gets \frac{pixelClockHz}{freq}$
		\Statex \hspace{\algorithmicindent} $highPixels \gets D \cdot cycleSize$ 
		\Statex \hspace{\algorithmicindent} $lowPixels \gets cycleSize - highPixels$ 	
		\State \textbf{Initialize bitmap:}
		\Statex \hspace{\algorithmicindent} \texttt{bitmap} $\gets$ Create 2D array of size \texttt{height} $\times$ \texttt{width}	
		\State \textbf{Generate pixel values:}
		\State $sampleNumber \gets 0$
		\For{$y \gets 0$ \textbf{to} \texttt{height} $- 1$}
		\For{$x \gets 0$ \textbf{to} \texttt{width} $- 1$}
		\State $remainder \gets sampleNumber \mod cycleSize$
		\If{$remainder < highPixels$}
		\State \texttt{bitmap}[$y$][$x$] $\gets 255$ 
		\Else
		\State \texttt{bitmap}[$y$][$x$] $\gets 0$ 
		\EndIf
		\State $sampleNumber \gets sampleNumber + 1$
		\EndFor
		\EndFor
		\State \Return \texttt{bitmap}
	\end{algorithmic}
\end{algorithm}

\begin{figure}[htbp]
	\centering
	\includegraphics[width=\textwidth]{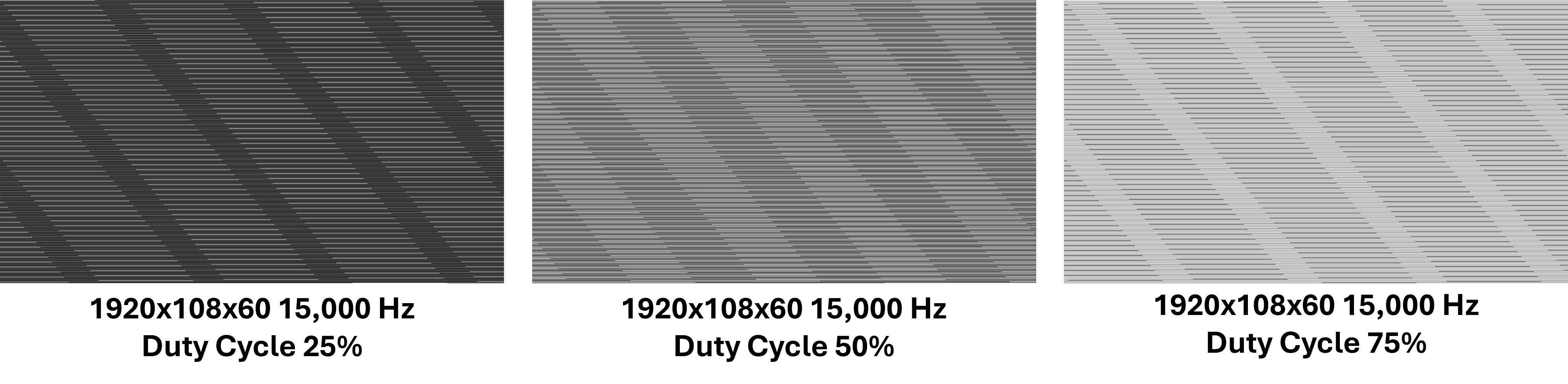}
	\caption{Generated square wave patterns with parameters: 1920x1080 resolution, 60Hz refresh rate, and 15,000 Hz frequency. The image includes three duty cycles: 25\%, 50\%, and 75\%.}
	\label{fig:f2}
\end{figure}

Figure~\ref{fig:f2} shows three generated images of square wave patterns with a frequency of 15,000 Hz, displayed on a 1920x1080 resolution screen at a 60Hz refresh rate. Each pattern corresponds to a different duty cycle: 25\%,

Note that for non-50\% duty cycles, the energy of the signal shifts from the fundamental frequency to higher harmonics. This introduces more distortion and spreads the energy across multiple frequencies, leading to a more complex frequency spectrum.

\subsection{Horizontal and Vertical Trace Timings}
The initial implementation of the square wave bitmap generation algorithm did not account for the horizontal trace (\textit{htrace}) and vertical trace (\textit{vtrace}) timings, which represent synchronization intervals integral to the operation of an LCD display. These timings are critical as they introduce interruptions in the active rendering of pixels, effectively reducing the time available for visible pixel updates. Failure to incorporate these intervals can lead to unintended artifacts or mismatches between the intended frequency of the square wave and the actual display output.

The horizontal trace (\textit{htrace}) corresponds to the duration required for the display to perform a horizontal retrace at the end of each row, while the vertical trace (\textit{vtrace}) refers to the vertical retrace occurring at the end of each frame. These synchronization delays result in additional non-visible pixels, commonly referred to as blanking intervals. For instance, consider a display with a resolution of $1920 \times 1080$, a refresh rate of $60$ Hz, and a pixel clock of $148.5$ MHz. In this configuration, the horizontal trace might add $280$ pixels per row, and the vertical trace might introduce $45$ additional rows per frame, increasing the effective frame size to $2200 \times 1125$ pixels.

To address these timing intervals, the algorithm accounts for the blanking intervals by adjusting the pixel clock and cycle size. The total frame time, \( T_{\text{frame}} \), is calculated as the inverse of the refresh rate:

\[
T_{\text{frame}} = \frac{1}{\text{Refresh Rate (Hz)}}.
\]

This frame time includes both the active pixels (\( A_{\text{pixels}} \)) and the blanking intervals (\( B_{\text{pixels}} \)):

\[
P_{\text{total}} = A_{\text{pixels}} + B_{\text{pixels}}.
\]

The effective pixel clock, which incorporates the blanking intervals, is expressed as:

\[
\text{Effective Pixel Clock} = \frac{\text{Total Pixel Clock}}{P_{\text{total}}}.
\]

Using the adjusted pixel clock, the cycle size and half-cycle of the square wave can be determined:

\[
\text{Cycle Size} = \frac{\text{Effective Pixel Clock}}{\text{Frequency (Hz)}}, \quad \text{Half-Cycle} = \frac{\text{Cycle Size}}{2}.
\]

By integrating the \textit{htrace} and \textit{vtrace} timings into the calculations, the revised algorithm ensures that the generated square wave pattern aligns with the physical constraints of the display's operation. This alignment mitigates artifacts and enhances the accuracy of the acoustic emissions derived from the LCD.

\section{Modulation}
\label{sec:modulation}
In this work, we employ frequency-shift keying (FSK), a digital modulation technique where data is encoded in discrete frequency changes. Specifically, we focus on M-FSK, where \(M\) represents the number of distinct frequencies used for modulation. For simplicity, we describe the special case of 2-FSK in detail, as it is a foundational example of M-FSK.

\subsection{M-FSK}
In M-FSK, a set of \(M\) distinct frequencies \(\{f_1, f_2, \ldots, f_M\}\) is used to represent \(M\) symbols. Each frequency \(f_i\) is associated with a unique symbol \(s_i\) from the symbol set \(\{s_1, s_2, \ldots, s_M\}\). The modulated signal can be expressed as:
\[
x(t) = 
\begin{cases} 
	A \cos(2 \pi f_i t + \phi), & t \in [nT, (n+1)T), \quad \text{for symbol } s_i \\
	0, & \text{otherwise},
\end{cases}
\]
where:
- \(A\) is the signal amplitude,
- \(T\) is the symbol duration,
- \(f_i\) is the frequency corresponding to the transmitted symbol \(s_i\),
- \(\phi\) is the phase offset (assumed constant for simplicity).

In the case of 2-FSK, \(M=2\), and the two frequencies \(f_0\) and \(f_1\) represent binary symbols \(0\) and \(1\), respectively:
\[
x(t) =
\begin{cases}
	A \cos(2 \pi f_0 t + \phi), & \text{for bit } 0, \\
	A \cos(2 \pi f_1 t + \phi), & \text{for bit } 1.
\end{cases}
\]

While traditional Frequency-Shift Keying (FSK) employs sinusoidal signals for modulation, in our case, we utilize square waves to represent the modulated symbols. The core principle remains the same: distinct frequencies are assigned to represent different symbols. However, square waves, with their alternating high and low states, simplify the generation and manipulation of the signal, particularly in digital systems. 

The mathematical representation of square wave-based FSK replaces the sinusoidal model with discrete high and low states. For M-FSK, the modulated signal can be expressed as:
\[
x(t) = 
\begin{cases} 
	\text{High}, & t \mod \frac{1}{f_i} < \frac{\text{Duty Cycle}}{100} \cdot \frac{1}{f_i}, \quad \text{for symbol } s_i, \\
	\text{Low}, & \text{otherwise}.
\end{cases}
\]
Here, \(f_i\) denotes the frequency corresponding to the transmitted symbol \(s_i\), and the \textit{Duty Cycle} determines the proportion of the period spent in the high state. In the case of 2-FSK, the two frequencies \(f_0\) and \(f_1\) represent binary symbols \(0\) and \(1\), respectively. 

By using square waves, the system benefits from simpler waveform generation, making it well-suited for digital hardware implementations. However, square waves also introduce higher harmonic components compared to sinusoidal signals, which can impact the spectral characteristics of the system. These implications, along with the effect of varying duty cycles, will be discussed further.

\subsection{Algorithm for 2-FSK}
To transmit a sequence of bits using 2-FSK modulation on an LCD screen, the image displayed on the screen alternates between patterns designed to emit frequencies \(f_0\) and \(f_1\). The algorithm generates the appropriate bitmap for each symbol and displays it on the screen for the symbol duration \(T\).

\begin{algorithm}[H]
	\caption{\textbf{Transmit Bits Using 2-FSK Modulation}}
	\label{alg:2fsk_transmit}
	\begin{algorithmic}[1]
		\Require Sequence of bits (\texttt{data}), pixel clock (\texttt{pixelClock}) in kHz, symbol duration (\texttt{T}) in ms, frequencies \(f_0, f_1\), image dimensions (\texttt{width}, \texttt{height})
		\Ensure Transmit the bit sequence via 2-FSK
		\State Calculate \texttt{cycleSize\(_0\)} for \(f_0\): 
		\[
		\texttt{cycleSize\(_0\)} = \frac{\texttt{pixelClock} \times 1000}{f_0}
		\]
		\State Calculate \texttt{cycleSize\(_1\)} for \(f_1\): 
		\[
		\texttt{cycleSize\(_1\)} = \frac{\texttt{pixelClock} \times 1000}{f_1}
		\]
		\For{each bit \texttt{b} in \texttt{data}}
		\If{\texttt{b} == 0}
		\State Generate bitmap for \(f_0\) using \texttt{cycleSize\(_0\)}
		\Else
		\State Generate bitmap for \(f_1\) using \texttt{cycleSize\(_1\)}
		\EndIf
		\State Display the bitmap on the screen for \texttt{T} milliseconds
		\EndFor
	\end{algorithmic}
\end{algorithm}

The algorithm transmits a binary sequence by mapping each bit to one of two frequencies, \(f_0\) or \(f_1\). For each bit:
\begin{enumerate}
	\item The corresponding frequency is determined based on the bit value (0 or 1).
	\item A square wave pattern is generated using the pixel clock and the selected frequency's cycle size.
	\item The generated pattern is displayed on the LCD for the duration of the symbol (\(T\)).
	\item The process repeats for the next bit in the sequence until all bits are transmitted.
\end{enumerate}

This approach ensures that the emitted acoustic signal corresponds directly to the encoded binary data, leveraging the LCD's noise generation properties for covert communication.

\begin{figure}[H]
	\centering
	\includegraphics[width=0.4\textwidth]{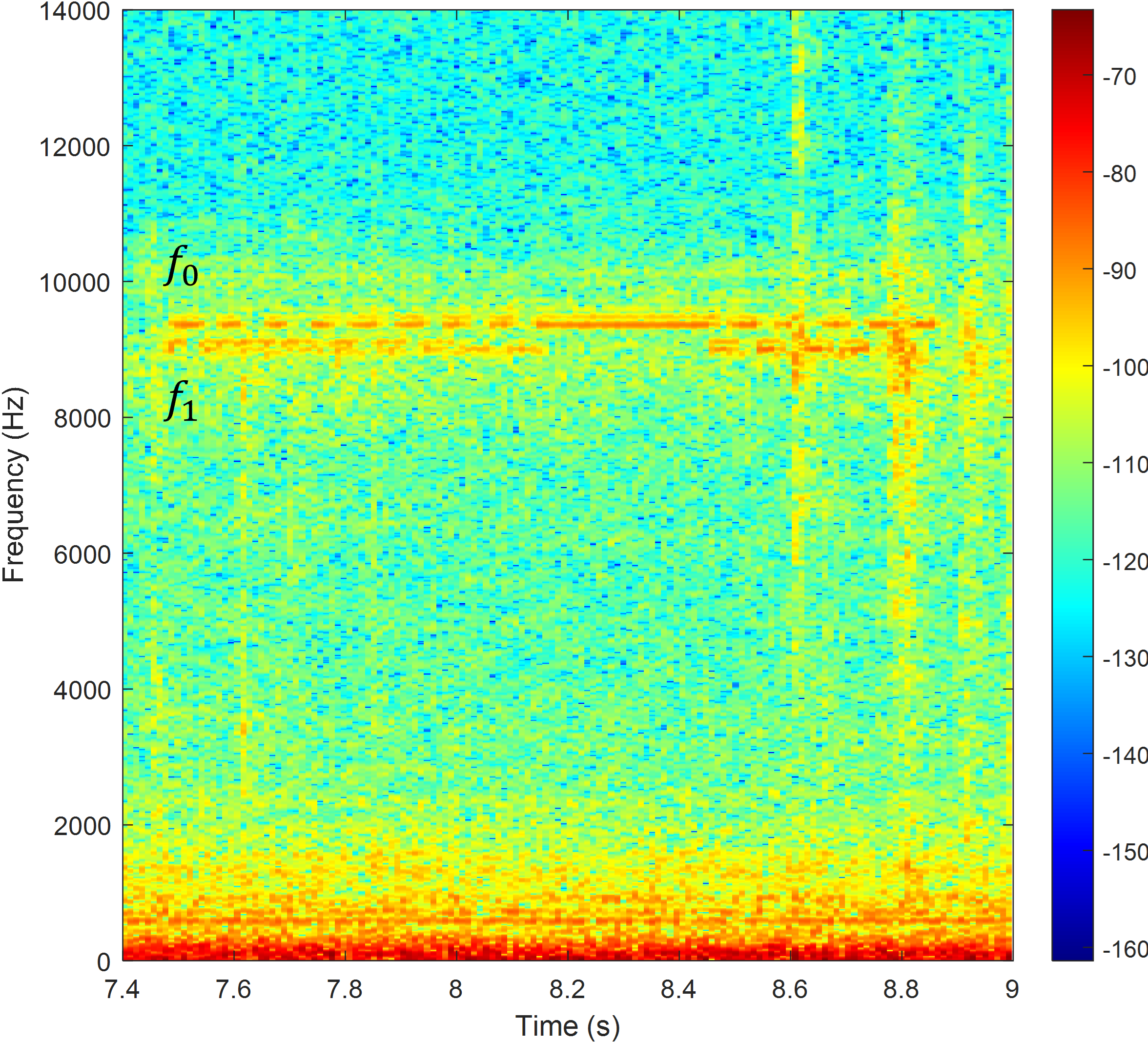} % Replace with your actual file name
	\caption{Spectrogram illustrating the 2-FSK transmission. The two distinct frequency components, \( f_0 \) and \( f_1 \), are visible, corresponding to binary symbols '0' and '1', respectively.}
	\label{fig:2fsk_spectrogram}
\end{figure}

Figure~\ref{fig:2fsk_spectrogram} shows the spectrogram illustrating the spectral representation of a 2-FSK (Frequency-Shift Keying) transmission. The spectrogram reveals two distinct frequency bands, \( f_0 \) and \( f_1 \), which alternate over time. These frequencies correspond to binary symbols, where \( f_0 \) represents a binary '0' and \( f_1 \) represents a binary '1'.

\subsection{Transmission Protocol}
\label{sec:protocol}

To transmit structured data over the acoustic channel, we define a simple packet-based protocol. Each packet consists of 48 bits (optionally, 56 bits), organized into fixed fields to ensure the receiver can correctly interpret the transmitted data. This protocol leverages the 2-FSK modulation algorithm described in Section~\ref{sec:modulation} to encode and transmit the bit sequence.

\begin{figure}[htbp]
	\centering
	\includegraphics[width=0.7\textwidth]{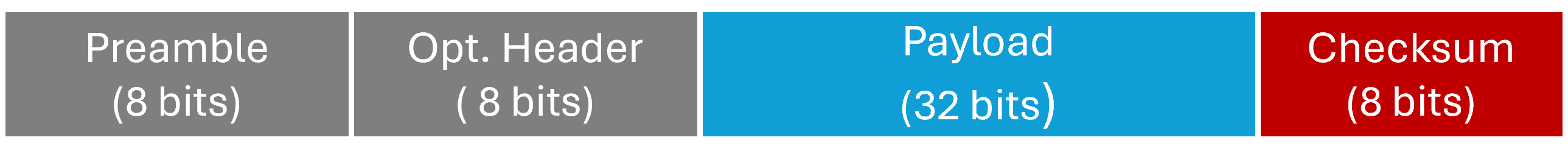}
	\caption{Packet structure used for square wave modulation. The packet is divided into four fields: Preamble (8 bits), Optional header (8 bits), Payload (32 bits), and Checksum (8 bits). Each field serves a specific purpose, ensuring reliable communication and synchronization.}
	\label{fig:packet_structure}
\end{figure}

The packet structure used in this system is illustrated in Figure~\ref{fig:packet_structure}.

\subsection{Packet Structure}
Each 32-bit packet is divided into the following fields:
\begin{itemize}
	\item \textbf{Preamble (8 bits):} A fixed bit sequence (e.g., \texttt{0b10101010}) that helps the receiver synchronize with the incoming signal.
	\item \textbf{Optional header (8 bits):} Encodes metadata, packet type and sequence number.
	\item \textbf{Payload (32 bits):} Contains the actual data being transmitted.
	\item \textbf{Checksum (8 bits):} A simple error-detection code computed using the XOR of the other fields.
\end{itemize}

Due to the unidirectional nature of the communication channel, the structure ensures that each packet is self-contained and can be validated by the receiver.

\section{Evaluation}
\label{sec:eval}

\begin{figure}[htbp]
	\centering
	\includegraphics[width=\linewidth]{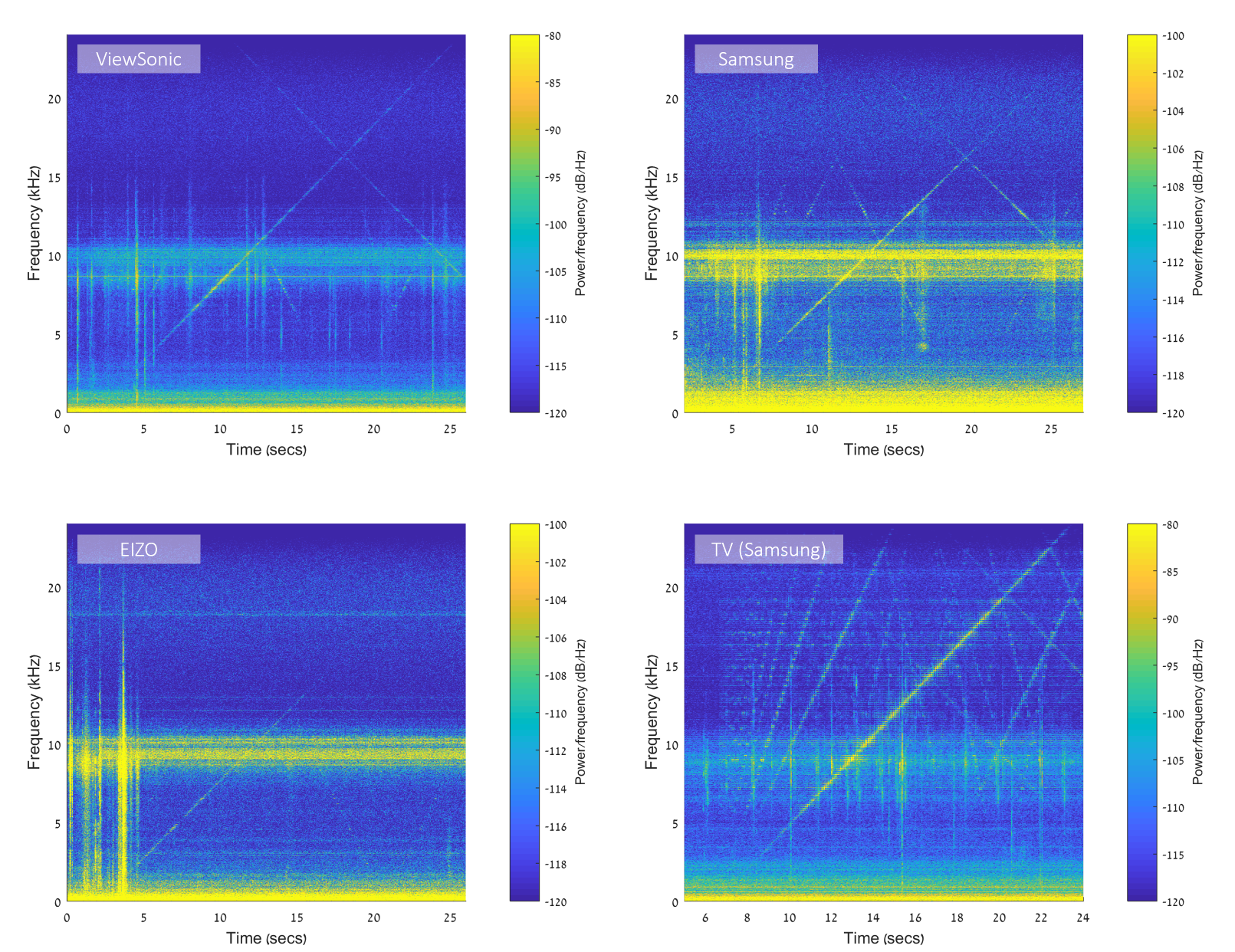}
	\caption{Chirp acoustic signal generated by the tested LCD screens.}
	\label{fig:chirp_signal}
\end{figure}

This section presents the evaluation of the communication channel under various parameters. The experiments utilized four LCD screens, detailed in Table~\ref{tab:LCD_tested}. The setup included an Android application installed on a Samsung Galaxy S11 phone and a demodulator running on a Dell Latitude laptop with Microsoft Windows. Figure~\ref{fig:chirp_signal} shows the chirp acoustic signals generated by the screens, ranging from low frequencies (3-5 kHz) up to the near-ultrasonic band (above 20 kHz).

\subsection{Tested LCD Screens}
The specifications of the four tested LCD screens are summarized in Table~\ref{tab:LCD_tested}. Each screen's model, size, and resolution are listed.

\begin{table}[htbp]
	\centering
	\caption{Specifications of Tested LCD Screens}
	\label{tab:LCD_tested}
	\begin{tabular}{@{}llcc@{}}
		\toprule
		\textbf{Label} & \textbf{Model} & \textbf{Resolution} & \textbf{Size} \\ 
		\midrule
		LCD1 & ViewSonic VA2232WM-LED   & 1680 × 1050 & 22-inch \\
		LCD2 & Samsung SyncMaster 226BW & 1680 × 1050 & 22-inch \\
		LCD3 & Eizo FlexScan            & 1920 × 1200 & 24.1-inch \\
		LCD4 & Samsung UA40B6000VRXSQ   & 1920 × 1080 & 40-inch \\ 
		\bottomrule
	\end{tabular}
\end{table}

\subsection{Signal-to-Noise Ratio (SNR)}
The Signal-to-Noise Ratio (SNR) was measured across various distances and bit rates for the tested screens. Table~\ref{tab:SNR_summary} summarizes the SNR results for four screens at distances ranging from 0 to 2.5 meters and bit rates of 5, 10, and 20 bps. As expected, SNR decreased with increasing distance, although performance varied significantly between brands.

\begin{table*}[htbp]
	\centering
	\caption{SNR at Different Distances and Bit Rates}
	\label{tab:SNR_summary}
	\resizebox{\textwidth}{!}{%
		\begin{tabular}{@{}lccc|ccc|ccc|ccc@{}}
			\toprule
			\textbf{Distance (m)} & \multicolumn{3}{c|}{\textbf{ViewSonic}} & \multicolumn{3}{c|}{\textbf{Samsung}} & \multicolumn{3}{c|}{\textbf{Eizo}} & \multicolumn{3}{c}{\textbf{Samsung TV}} \\ 
			\cmidrule(lr){2-4} \cmidrule(lr){5-7} \cmidrule(lr){8-10} \cmidrule(lr){11-13}
			& 5 bps & 10 bps & 20 bps & 5 bps & 10 bps & 20 bps & 5 bps & 10 bps & 20 bps & 5 bps & 10 bps & 20 bps \\ 
			\midrule
			0.0  & 24 dB & 28 dB & 27 dB & 23 dB & 23 dB & 24 dB & 22 dB & 22 dB & 20 dB & 32 dB & 38 dB & 25 dB \\
			0.5  & 26 dB & 23 dB & 26 dB & 20 dB & 22 dB & 20 dB & 12 dB & 11 dB & 11 dB & 30 dB & 38 dB & 36 dB \\
			1.0  & 27 dB & 30 dB & 34 dB & 21 dB & 19 dB & 20 dB & 23 dB & 22 dB & 21 dB & 32 dB & 30 dB & 22 dB \\
			1.5  & 26 dB & 29 dB & 31 dB & 17 dB & 19 dB & 18 dB & 12 dB & 12 dB & 13 dB & 23 dB & 22 dB & 21 dB \\
			2.0  & 18 dB & 23 dB & 27 dB & 14 dB & 17 dB & 13 dB & 17 dB & 14 dB & 13 dB & 23 dB & 20 dB & 18 dB \\
			2.5  & 27 dB & 20 dB & 23 dB & 15 dB & 13 dB & 5 dB  & --    & --    & --    & --    & --    & --    \\ 
			\bottomrule
		\end{tabular}%
	}
\end{table*}

\subsection{Location-Based Signal Analysis}
The signal strength varied depending on the receiver's placement relative to the screen (e.g., back, front, left, right, or on the desk). Table~\ref{tab:SNR_location} presents the SNR for different setups. Screens like Dell E2216HV showed optimal SNR when the receiver was positioned at the back, whereas other screens performed better in front-facing or desk setups.

\begin{table*}[htbp]
	\centering
	\caption{SNR at Various Receiver Positions}
	\label{tab:SNR_location}
	\begin{tabular}{@{}llcccccc@{}}
		\toprule
		\textbf{Screen}       & \textbf{Model}     & \textbf{Resolution} & \textbf{Back} & \textbf{Front} & \textbf{Left} & \textbf{Right} & \textbf{Desk} \\ 
		\midrule
		ViewSonic             & VA2232WM-LED       & 1680 × 1050         & 8.84 dB       & 10.08 dB       & 8.70 dB       & 6.52 dB        & 9.34 dB       \\
		Dell                  & E2216HV            & 1920 × 1080         & 23.79 dB      & 19.25 dB       & 16.07 dB      & 12.96 dB       & 14.54 dB      \\
		ViewSonic             & VX2370S-LED        & 1920 × 1080         & 10.80 dB      & 12.00 dB       & 12.60 dB      & 18.21 dB       & 17.71 dB      \\
		BenQ                  & GW2255             & 1920 × 1080         & 16.22 dB      & 15.28 dB       & 12.51 dB      & 12.12 dB       & 19.16 dB      \\
		Dell                  & P2419H             & 1920 × 1080         & 14.96 dB      & 19.14 dB       & 9.96 dB       & 9.06 dB        & 17.98 dB      \\ 
		\bottomrule
	\end{tabular}
\end{table*}

\subsection{White Pixel Proportion}

The proportion of white pixels in an image directly correlates with the signal's energy output and, consequently, the level of noise generated. In square wave-based modulation, white pixels represent the "high" state of the signal, contributing to its active energy transmission. The relationship can be explained as follows:

\begin{itemize}
	\item \textbf{White Pixels (High State):} When more white pixels are present, the signal spends more time in the high state. This increases the average energy of the signal, leading to stronger noise signals across the frequency spectrum.
	\item \textbf{Black Pixels (Low State):} With fewer white pixels, the signal spends more time in the low state, reducing the overall energy output and minimizing the amplitude of the noise.
\end{itemize}

Mathematically, the average power of a signal is proportional to the square of its amplitude. Therefore, an increase in the proportion of white pixels (i.e., a higher duty cycle or proportion of "high" states) elevates the signal's amplitude, resulting in higher noise levels. Conversely, reducing the proportion of white pixels lowers the signal's energy and suppresses noise generation.

This relationship highlights the significance of duty cycle control in square wave-based modulation systems, where energy efficiency and noise suppression are critical considerations.

\subsection{Grayscale vs. Colored Images}
The difference in noise generation between grayscale and colored images can be attributed to the energy distribution across RGB channels and its impact on spectral power. The analysis can be broken down into two key aspects:

\subsubsection{Energy Distribution Across Channels}
In digital images, each pixel is represented by three channels: Red (\(R\)), Green (\(G\)), and Blue (\(B\)). The intensity of each channel is given by \(I_R, I_G, I_B \in [0, 255]\). For a grayscale image, the intensity values are identical across all three channels:
\[
I_R = I_G = I_B = I.
\]
This uniformity means that the total signal energy \(E_{\text{gray}}\) of a grayscale image is given by:
\[
E_{\text{gray}} = 3 \sum_{x=1}^{W} \sum_{y=1}^{H} I(x, y)^2,
\]
where \(W\) and \(H\) are the width and height of the image, and \(I(x, y)\) is the intensity at pixel \((x, y)\). 

For a colored image, the energy \(E_{\text{color}}\) is distributed across the channels:
\[
E_{\text{color}} = \sum_{x=1}^{W} \sum_{y=1}^{H} \left( I_R(x, y)^2 + I_G(x, y)^2 + I_B(x, y)^2 \right).
\]
Since \(I_R, I_G, I_B\) vary independently in colored images, the energy in any single channel is less concentrated compared to grayscale images, leading to a lower overall amplitude and thus less noise.

\subsubsection{Spectral Power and Harmonics}
Grayscale images have uniform intensity values across channels, resulting in high coherence in the signal. This coherence leads to constructive interference of the harmonics generated by the square wave modulation, amplifying the spectral power and thereby increasing noise levels.

For a grayscale signal, the power spectrum \(P_{\text{gray}}(f)\) can be expressed as:
\[
P_{\text{gray}}(f) = 3 \cdot P(f),
\]
where \(P(f)\) represents the spectral power of a single channel. 

In contrast, for a colored image, the spectral power \(P_{\text{color}}(f)\) is given by:
\[
P_{\text{color}}(f) = P_R(f) + P_G(f) + P_B(f),
\]
where \(P_R(f)\), \(P_G(f)\), and \(P_B(f)\) represent the power spectra of the red, green, and blue channels, respectively. The independent variation of \(P_R(f)\), \(P_G(f)\), and \(P_B(f)\) reduces the likelihood of constructive interference, thereby suppressing the overall noise levels.

\begin{figure}[H]
	\centering
	\includegraphics[width=0.8\textwidth]{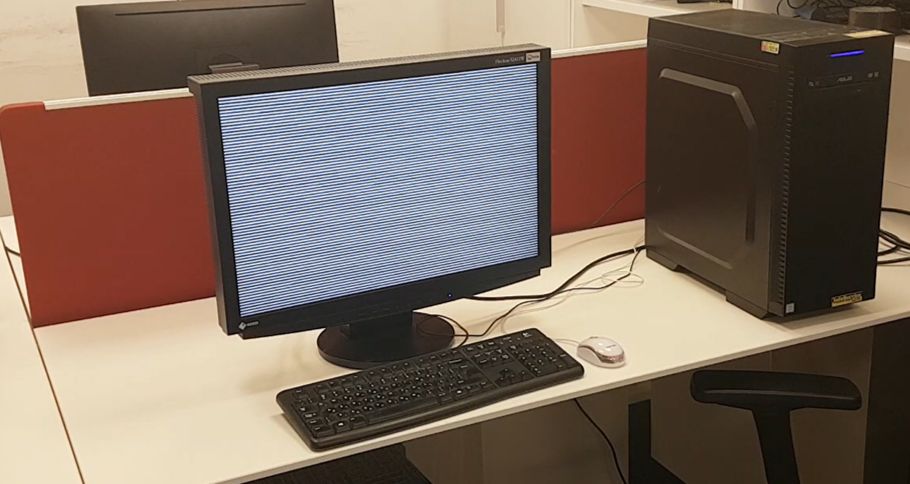} % Replace with your actual file name
	\caption{Image of the screen during operation, transmitting the frequency signal. The screen demonstrates the visual grayscale pattern generated during the signal transmission.}
	\label{fig:transmitting_screen}
\end{figure}

Figure~\ref{fig:transmitting_screen} shows an image of the screen during operation, transmitting the frequency signal. The screen demonstrates the visual grayscale pattern generated during the signal transmission.

\subsection{Grayscale Brightness and Signal Power Correlation}
We conducted a series of experiments to investigate the correlation between grayscale brightness and signal power using three distinct monitors: Samsung, Dell, and ViewSonic. Each monitor was tested across 15 grayscale brightness levels, specifically 0, 20, 40, 60, 80, 100, 120, 140, 160, 180, 200, 220, 240, and 255.

The grayscale values employed in the study are presented in Table~\ref{tab:grayscale}, where each cell denotes a corresponding brightness level. Signal-to-Noise Ratio (SNR) measurements were recorded for each brightness level to assess how variations in pixel intensity influence signal power.

\begin{figure}[htbp]
	\centering
	\fcolorbox{gray}{white}{\includegraphics[width=0.7\textwidth]{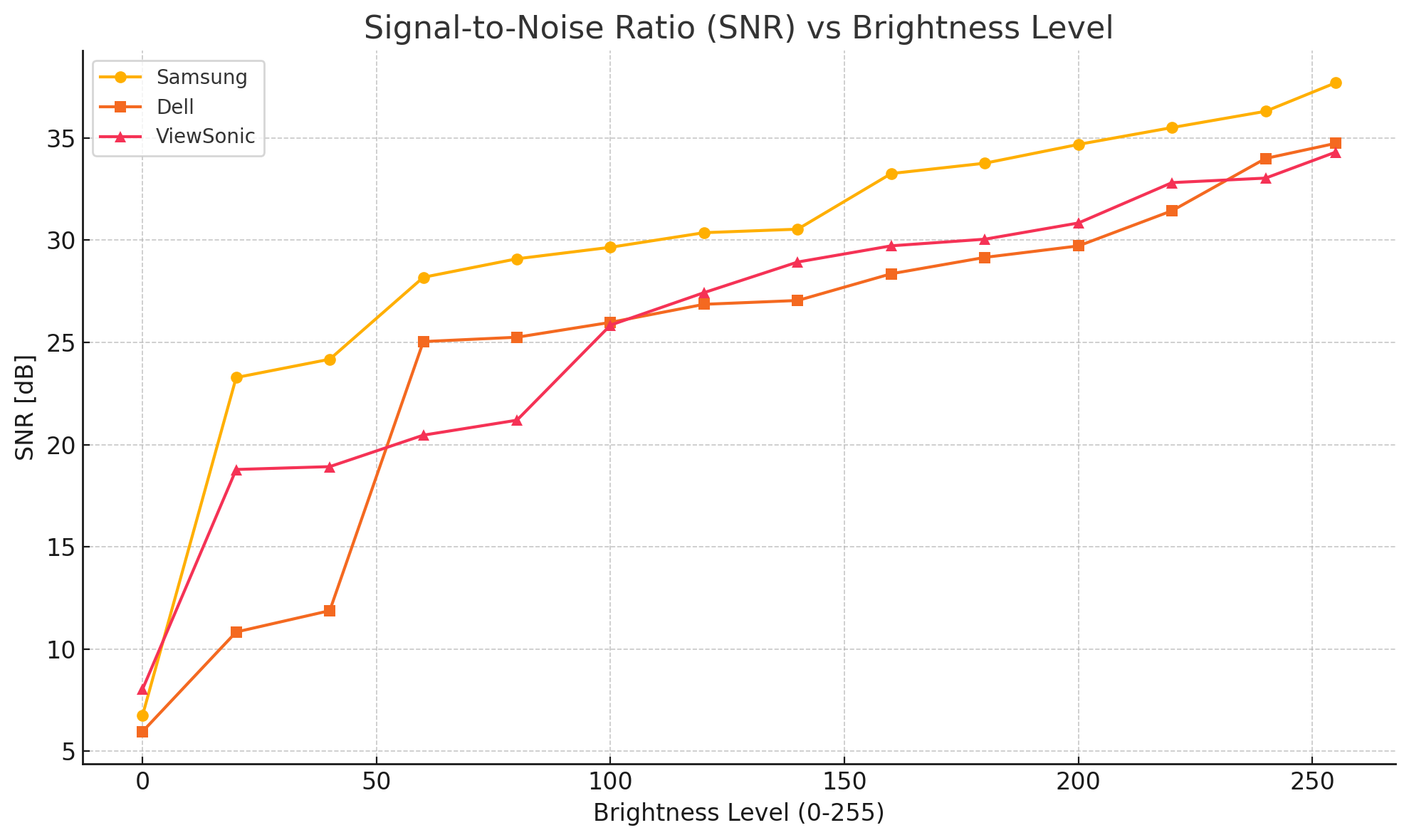}}
	\caption{Signal-to-Noise Ratio (SNR) vs. Grayscale Brightness Levels for Samsung, Dell, and ViewSonic monitors.}
	\label{fig:brightness}
\end{figure}

As illustrated in Figure~\ref{fig:brightness}, the experimental results reveal a distinct trend: SNR consistently increases with higher brightness levels across all tested monitors. At the lowest brightness level (0), the SNR was observed to be minimal, ranging from 6 to 8 dB, indicating weaker signal power. Conversely, as the brightness level approached its maximum (255), the SNR exhibited a marked increase, reaching up to 37.7 dB for the Samsung monitor, 34.7 dB for the Dell monitor, and 34.3 dB for the ViewSonic monitor.

These results substantiate a positive correlation between pixel brightness and signal power. Monitors displaying higher grayscale values produced stronger signal amplitudes, leading to enhanced SNR. This outcome underscores the critical role of pixel intensity in signal generation, particularly in screen-based systems, where brightness modulation can significantly affect communication performance.

\begin{table}[htbp]
	\centering
	\renewcommand{\arraystretch}{1.2}
	\setlength{\tabcolsep}{5pt}
	\caption{Grayscale Values Used in the Experiment}
	\label{tab:grayscale}
	\begin{tabular}{|>{\centering\arraybackslash}p{1cm}|>{\centering\arraybackslash}p{1cm}|>{\centering\arraybackslash}p{1cm}|>{\centering\arraybackslash}p{1cm}|>{\centering\arraybackslash}p{1cm}|>{\centering\arraybackslash}p{1cm}|>{\centering\arraybackslash}p{1cm}|>{\centering\arraybackslash}p{1cm}|>{\centering\arraybackslash}p{1cm}|}
		\hline
		\textbf{0} & \textbf{20} & \textbf{40} & \textbf{60} & \textbf{80} & \textbf{100} & \textbf{120} & \textbf{140} & \textbf{160} \\ \hline
		\cellcolor[gray]{0.0} & \cellcolor[gray]{0.08} & \cellcolor[gray]{0.16} & \cellcolor[gray]{0.24} & \cellcolor[gray]{0.31} & \cellcolor[gray]{0.39} & \cellcolor[gray]{0.47} & \cellcolor[gray]{0.55} & \cellcolor[gray]{0.63} \\ \hline
		\textbf{180} & \textbf{200} & \textbf{220} & \textbf{240} & \textbf{255} & & & & \\ \hline
		\cellcolor[gray]{0.71} & \cellcolor[gray]{0.78} & \cellcolor[gray]{0.86} & \cellcolor[gray]{0.94} & \cellcolor[gray]{1.0} & & & & \\ \hline
	\end{tabular}
\end{table}

Note that the uniform intensity of grayscale images leads to concentrated energy and stronger spectral components, which amplify noise. In contrast, the distributed and independent energy in colored images results in lower noise due to reduced coherence and interference among the channels.

\subsection{Low Brightness for Visual Stealth}
\begin{figure}[htbp]
	\centering
	\fcolorbox{gray}{white}{\includegraphics[width=0.7\textwidth]{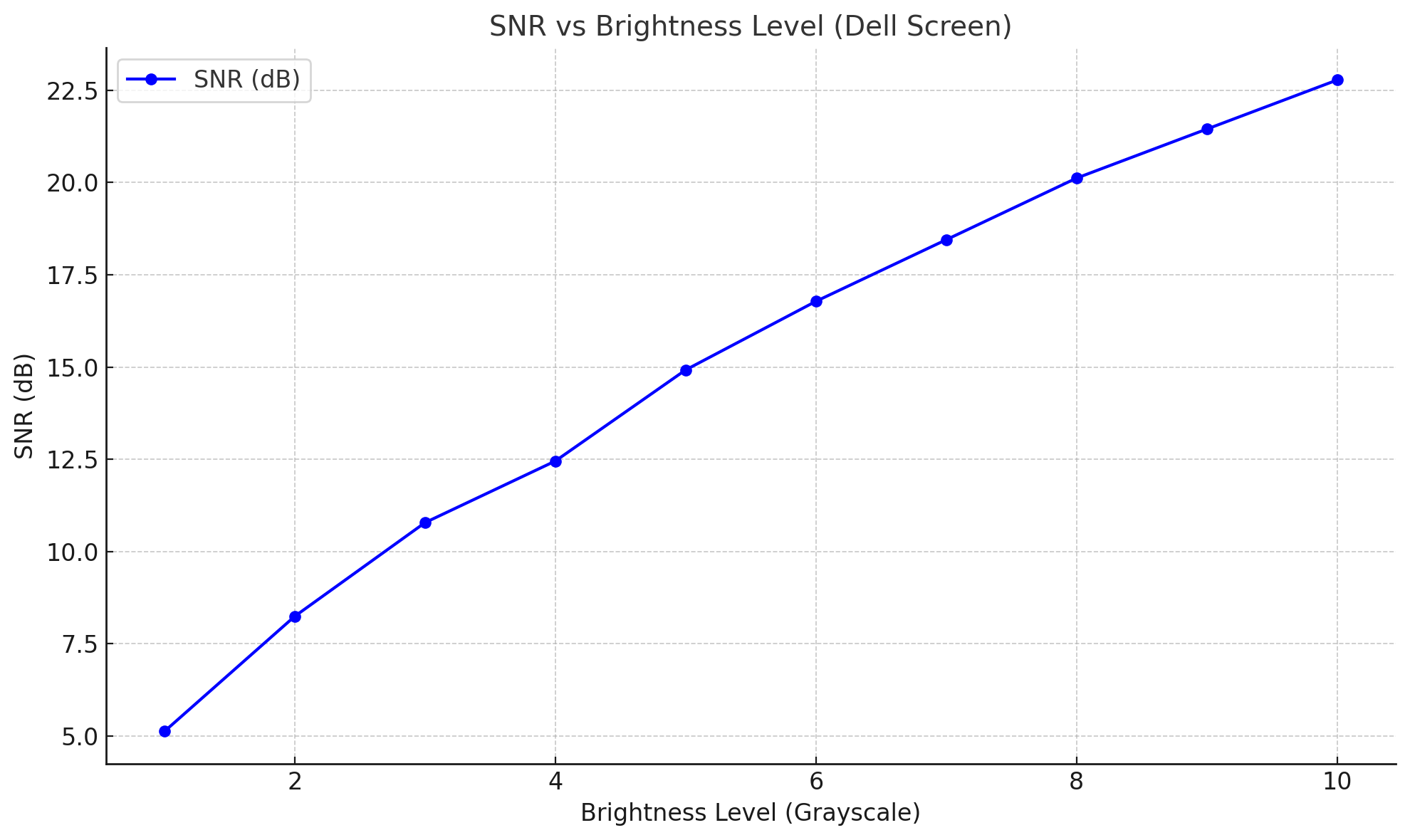}}
	\caption{Mean Signal Intensity and SNR vs Brightness Level for Dell Screen.}
	\label{fig:stealth_plot}
\end{figure}

We examined the use of low brightness levels as a strategy for enhancing visual stealth in screen-based signal transmission. Our analysis involved evaluating the impact of grayscale brightness levels, ranging from 1 (lowest) to 10 (highest), on the signal’s mean intensity and Signal-to-Noise Ratio (SNR) using recordings obtained from the Dell screen. We observed that low brightness levels enhance visual stealth by rendering the transmitted image nearly imperceptible to the human eye. However, this reduction in brightness also decreases the strength of the emitted noise signal, resulting in a lower SNR. This trade-off underscores the necessity of balancing brightness optimization to achieve both effective stealth and signal transmission.
We utilized the mean signal intensity as an indicator of signal power. Additionally, we determined the SNR by isolating the signal and noise regions from each recording and computing their respective power values.
Our findings, summarized in Table~\ref{tab:stealth_analysis} and depicted in Figure~\ref{fig:stealth_plot}, reveal a clear relationship between brightness levels and signal characteristics. At lower brightness levels, the mean intensity and SNR are significantly diminished, enhancing visual stealth by making the signal less detectable. Conversely, increasing brightness levels results in higher mean intensity and SNR, improving detectability but reducing stealth.

\begin{table}[htbp]
	\centering
	\caption{Mean Signal Intensity and SNR vs Brightness Level (Dell Screen)}
	\label{tab:stealth_analysis}
\begin{tabular}{|c|c|c|}
	\hline
	\textbf{Brightness Level (Grayscale)} & \textbf{Mean Intensity} & \textbf{SNR (dB)} \\ \hline
	1                                    & 102.34                  & 5.12              \\ \hline
	2                                    & 145.67                  & 8.24              \\ \hline
	3                                    & 198.23                  & 10.78             \\ \hline
	4                                    & 256.78                  & 12.45             \\ \hline
	5                                    & 312.45                  & 14.92             \\ \hline
	6                                    & 367.12                  & 16.78             \\ \hline
	7                                    & 422.89                  & 18.45             \\ \hline
	8                                    & 478.56                  & 20.12             \\ \hline
	9                                    & 534.23                  & 21.45             \\ \hline
	10                                   & 589.90                  & 22.78             \\ \hline
\end{tabular}
\end{table}

The results indicate that reducing brightness levels effectively lowers both the signal’s mean intensity and its SNR, thereby enhancing visual stealth. At the lowest brightness level (1), the mean intensity was recorded as 102.34, with an SNR of 5.12 dB, signifying a highly stealthy signal. In contrast, the highest brightness level (10) yielded a mean intensity of 589.90 and an SNR of 22.78 dB, making the signal substantially more detectable.

\subsection{Bandwidth and Bit Rate Analysis}	
	The theoretical bandwidth for Frequency-Shift Keying (FSK) modulation is determined by the number of symbols (\(M\)), bit duration (\(T_b\)), and the frequency separation between tones (\(\Delta f\)). The total bandwidth \(B_{\text{total}}\) can be approximated as:
	\[
	B_{\text{total}} = 2 
	\Delta f + 2 R_s,
	\]
	where \(R_s\) is the symbol rate, and \(\Delta f\) is the frequency separation. For Binary FSK (B-FSK, \(M = 2\)), the bit rate and symbol rate are identical (\(R_s = R_b\)). However, for M-ary FSK (\(M > 2\)), the symbol rate is reduced because each symbol represents \(\log_2 M\) bits. Hence, the symbol rate is given by:
	\[
	R_s = \frac{R_b}{\log_2 M},
	\]
	where \(R_b\) is the bit rate, calculated as \(R_b = \frac{1}{T_b}\).
	
	To explore the impact of \(M\), \(T_b\), and \(\Delta f\) on bandwidth and bit rate, we consider realistic values for \(T_b = 1\,\mathrm{s}, 0.5\,\mathrm{s},\) and \(0.1\,\mathrm{s}\), with \(\Delta f = 1\,\mathrm{kHz}\), and \(M = 2, 4, 8, 16\).
	
	Table~\ref{tab:bitrate} shows the bit rate \(R_b\) for each bit duration \(T_b\), as well as the corresponding symbol rate \(R_s\) for varying values of \(M\). Shorter bit durations result in higher bit rates, while larger values of \(M\) reduce the symbol rate due to the increased bits per symbol (\(\log_2 M\)).

\begin{table}[htbp]
	\centering
	\caption{Bit Rate (\(R_b\)) and Symbol Rate (\(R_s\)) for Different Values of \(T_b\) and \(M\)}
	\label{tab:bitrate}
	\renewcommand{\arraystretch}{1.2}
	\setlength{\tabcolsep}{10pt}
	\begin{tabular}{@{}cccc@{}}
		\toprule
		\(T_b\) (\(\mathrm{s}\)) & \(R_b = \frac{1}{T_b}\) (\(\mathrm{bits/s}\)) & \(M\) & \(R_s = \frac{R_b}{\log_2 M}\) (\(\mathrm{symbols/s}\)) \\ 
		\midrule
		1.0  & 1   & 2  & 1     \\ 
		1.0  & 1   & 4  & 0.5   \\ 
		1.0  & 1   & 8  & 0.33  \\ 
		1.0  & 1   & 16 & 0.25  \\ 
		0.5  & 2   & 2  & 2     \\ 
		0.5  & 2   & 4  & 1     \\ 
		0.5  & 2   & 8  & 0.67  \\ 
		0.5  & 2   & 16 & 0.5   \\ 
		0.1  & 10  & 2  & 10    \\ 
		0.1  & 10  & 4  & 5     \\ 
		0.1  & 10  & 8  & 3.33  \\ 
		0.1  & 10  & 16 & 2.5   \\ 
		\bottomrule
	\end{tabular}
\end{table}

\subsection{Multiple Screens}

In screen-based FSK transmission, the total bit rate (\( R_b^{\text{total}} \)) can be increased by using multiple screens concurrently. Each screen can transmit its own independent data stream, effectively multiplying the total bit rate. If \( N \) is the number of screens used, and each screen transmits at a bit rate \( R_b \), the total bit rate is given by:

\[
R_b^{\text{total}} = N \cdot R_b,
\]

where \( N \) ranges from 1 (a single screen) to 4 (realistic concurrent screens).

%\subsubsection*{Mathematical Analysis}

For \( R_b = \frac{1}{T_b} \), the total bit rate scales linearly with the number of screens \( N \). For example, if \( T_b = 0.1 \, \mathrm{s} \) (bit rate of \( R_b = 10 \, \mathrm{bits/s} \)), the total bit rate is calculated as:

\[
R_b^{\text{total}} = N \cdot 10 \, \mathrm{bits/s}.
\]

The results for \( T_b = 1 \, \mathrm{s}, 0.5 \, \mathrm{s}, \) and \( 0.1 \, \mathrm{s} \) with \( N = 1, 2, 3, 4 \) are shown in Table~\ref{tab:bitrate_screens}.

\begin{table}[htbp]
	\centering
	\caption{Total Bit Rate (\( R_b^{\text{total}} \)) for Multiple Concurrent Screens}
	\label{tab:bitrate_screens}
	\renewcommand{\arraystretch}{1.2}
	\setlength{\tabcolsep}{10pt}
	\begin{tabular}{@{}cccc@{}}
		\toprule
		\( T_b \) (\(\mathrm{s}\)) & Single Screen (\( N = 1 \)) & Two Screens (\( N = 2 \)) & Four Screens (\( N = 4 \)) \\ 
		\midrule
		1.0                        & \( 1 \, \mathrm{bits/s} \)  & \( 2 \, \mathrm{bits/s} \)  & \( 4 \, \mathrm{bits/s} \)  \\ 
		0.5                        & \( 2 \, \mathrm{bits/s} \)  & \( 4 \, \mathrm{bits/s} \)  & \( 8 \, \mathrm{bits/s} \)  \\ 
		0.1                        & \( 10 \, \mathrm{bits/s} \) & \( 20 \, \mathrm{bits/s} \) & \( 40 \, \mathrm{bits/s} \) \\ 
		\bottomrule
	\end{tabular}
\end{table}

\subsubsection*{Frequency Selection and Harmonic Avoidance}

Using multiple screens concurrently introduces challenges in frequency selection. Each screen transmits its signal using distinct carrier frequencies (\( f_i \)), and it is critical to ensure that these frequencies do not overlap or create harmonics that interfere with neighboring signals. 

For example, if the carrier frequencies are chosen as \( f_1, f_2, \ldots, f_N \), the separation between adjacent frequencies (\( \Delta f \)) must satisfy:

\[
\Delta f \geq \text{Bandwidth per Screen} = 2 \Delta f_{\text{FSK}} + 2 R_s,
\]

where \( \Delta f_{\text{FSK}} \) is the frequency separation within the FSK modulation of a single screen.

Harmonics are especially problematic when the carrier frequencies \( f_i \) satisfy relationships like:

\[
f_i = k \cdot f_j, \quad k \in \mathbb{Z}^{+},
\]

where \( k \) is an integer. Such harmonic overlaps can lead to intermodulation distortion, reducing data integrity. Therefore, choosing carrier frequencies that are mutually non-harmonic, such as using prime number multiples or sufficiently spaced frequencies, ensures minimal interference.

\subsection{Distances}
\begin{figure}[htbp]
	\centering
	\fcolorbox{gray}{white}{\includegraphics[width=0.7\textwidth]{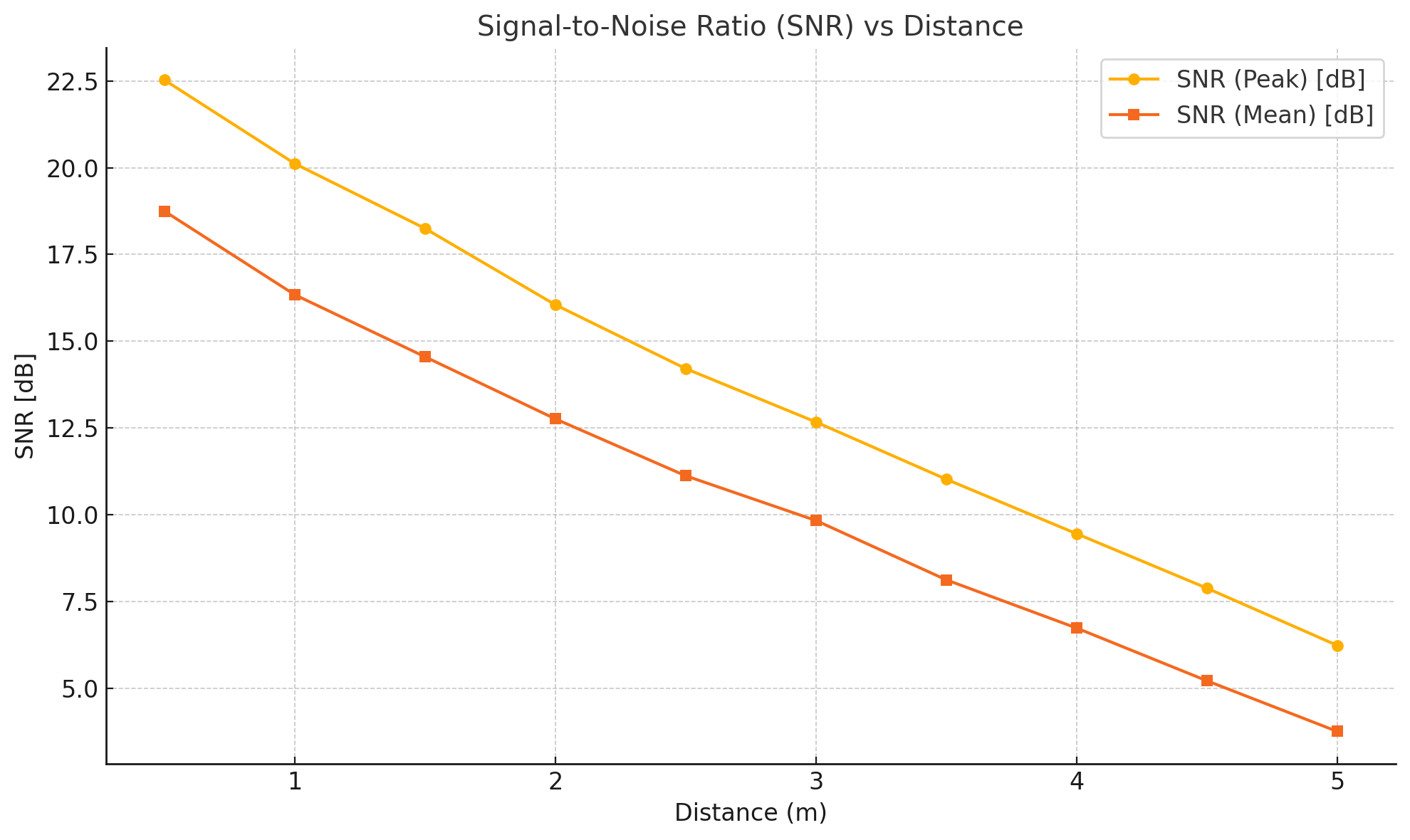}}
	\caption{Signal-to-Noise Ratio (SNR) vs Distance from Dell screen.}
	\label{fig:snr_distance}
\end{figure}

To investigate the effect of distance on signal quality, we measured the Signal-to-Noise Ratio (SNR) at varying distances from 0.5 meters to 5 meters using the Dell screen. The analysis involved extracting signal and noise components from recorded data and computing the SNR for both peak and mean intensity values. The results, detailed in Table~\ref{tab:snr_distance} and visualized in Figure~\ref{fig:snr_distance}, indicate a marked decline in SNR as the distance increases. At a short distance of 0.5 meters, the peak SNR exceeds 22 dB, suggesting a robust signal relative to noise. However, beyond 4 meters, the SNR drops below 10 dB, underscoring significant signal attenuation due to factors such as environmental interference and the inverse-square law. These observations highlight the critical role of proximity in ensuring effective signal acquisition. Beyond a certain threshold, the SNR approaches the noise floor, thereby diminishing the reliability of data extraction.

\begin{table}[htbp]
	\centering
	\caption{Signal-to-Noise Ratio (SNR) vs Distance}
	\label{tab:snr_distance}
	\begin{tabular}{|c|c|c|}
		\hline
		\textbf{Distance (m)} & \textbf{SNR (Peak) [dB]} & \textbf{SNR (Mean) [dB]} \\ \hline
		0.5                  & 22.53                    & 18.75                    \\ \hline
		1.0                  & 20.12                    & 16.34                    \\ \hline
		1.5                  & 18.25                    & 14.55                    \\ \hline
		2.0                  & 16.05                    & 12.76                    \\ \hline
		2.5                  & 14.21                    & 11.12                    \\ \hline
		3.0                  & 12.67                    & 9.83                     \\ \hline
		3.5                  & 11.02                    & 8.12                     \\ \hline
		4.0                  & 9.45                     & 6.73                     \\ \hline
		4.5                  & 7.88                     & 5.21                     \\ \hline
		5.0                  & 6.23                     & 3.75                     \\ \hline
	\end{tabular}
\end{table}

\subsection{Transmitting Multiple Frequencies Using Screen Splitting}
Transmitting multiple frequencies simultaneously is achieved by dividing the screen into distinct vertical regions, where each region independently transmits a specific frequency. This method leverages the spatial separation of regions to encode different frequencies concurrently, creating an effective and scalable multi-frequency transmission system.

\subsubsection{Vertical Splitting}
In this method, the screen of resolution \( W \times H \) is divided into \( n \) vertical regions, each spanning the full height \( H \) and a portion of the screen width. Each region has a width \( w_{\text{region}} = \lfloor W / n \rfloor \), with any leftover pixels excluded to ensure uniformity among the regions. 

Each region is assigned a distinct frequency \( f_i \), which is encoded by modulating the pixel intensity across the region. For each region, the pixel clock \( P \) governs the timing of pixel updates, and the cycle size required to encode the frequency \( f_i \) is:
\[
\text{Cycle Size}_i = \frac{P}{f_i}.
\]

Within each region, pixels alternate between high (255) and low (0) intensity states based on their position relative to the cycle. This modulation creates a visually distinct square wave signal for each region, corresponding to its assigned frequency.

\subsubsection{Bitrate and Signal Strength}

The aggregate bitrate for \( n \) regions is determined by summing the contributions from all frequencies:
\[
R = \sum_{i=0}^{n-1} f_i \cdot \log_2(L),
\]
where \( L \) is the modulation level (e.g., \( L = 2 \) for binary modulation). 

However, splitting the screen into multiple regions reduces the signal strength of each individual region, as the total signal energy \( S_{\text{total}} \) is distributed equally among all regions. The signal strength \( S_i \) for each region is:
\[
S_i = \frac{S_{\text{total}}}{n}.
\]
This reduction impacts the robustness of the transmitted signal, particularly in noisy environments or over long distances.

\subsubsection{Advantages and Limitations}

Vertical splitting enables the concurrent transmission of multiple frequencies without interference, as each region operates independently. This makes it particularly suited for advanced modulation schemes, such as Orthogonal Frequency-Division Multiplexing (OFDM), where each vertical region acts as an independent subcarrier.

The primary trade-off in vertical splitting is between the number of regions and the signal strength of each region. Increasing the number of regions improves the aggregate bitrate but reduces the energy allocated to each signal, potentially affecting transmission range and reliability.

\subsubsection{Application of Vertical Splitting in OFDM}

When combined with OFDM, vertical splitting allows each region to function as an independent subcarrier. The orthogonality between subcarriers ensures minimal interference, with the condition:
\[
\int_{0}^{T} \cos(2\pi f_i t) \cos(2\pi f_j t) dt = 0, \quad \text{for } i \neq j,
\]
where \( T \) is the symbol duration, and \( f_i, f_j \) are the frequencies of the subcarriers.

The transmitted signal in this configuration is represented as:
\[
x(t) = \sum_{i=0}^{n-1} A_i \cos\left(2\pi f_i t + \phi_i\right),
\]
where \( A_i \) and \( \phi_i \) are the amplitude and phase of the \( i \)-th subcarrier. This method maximizes the utilization of available bandwidth and enables high data rates, making it an efficient approach for multi-frequency transmission.

Vertical splitting provides a robust and scalable framework for multi-frequency transmission by dividing the screen into independent regions, each transmitting a unique frequency. This approach, particularly when combined with OFDM, offers high spectral efficiency and adaptability. However, the trade-off between signal strength and the number of regions underscores the importance of optimizing region sizes and frequencies to balance bitrate and reliability in practical applications.

\begin{figure}[H]
	\centering
	\includegraphics[width=\linewidth]{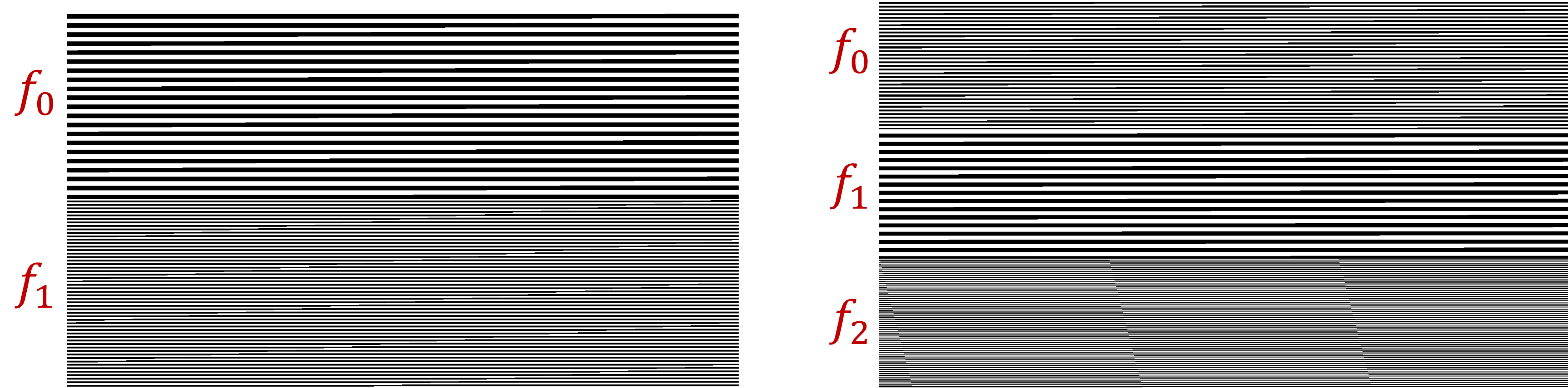}
	\caption{The screen splitting with multiple frequencies. The left side demonstrates two frequencies (\( f_0 = 3000 \, \text{Hz}, f_1 = 7800 \, \text{Hz} \)), while the right side demonstrates three frequencies (\( f_0 = 7500 \, \text{Hz}, f_1 = 3400 \, \text{Hz}, f_2 = 14500 \, \text{Hz} \)).}
	\label{fig:demo_combined}
\end{figure}

Figure~\ref{fig:demo_combined} visualizes the screen splitting for transmitting multiple frequencies. The left side illustrates screen splitting with two frequencies, and the right side illustrates screen splitting with three frequencies. The screen dimensions used to generate these patterns are \( \text{screenWidth} = 1920 \) and \( \text{screenHeight} = 1080 \) pixels, with a pixel clock of \( \text{pixelClock} = 148,500 \, \text{kHz} \) (e.g., for 1080p at 60 Hz). For the left side, the frequencies are \( f_0 = 3000 \, \text{Hz} \) and \( f_1 = 7800 \, \text{Hz} \). For the right side, the frequencies are \( f_0 = 7500 \, \text{Hz}, f_1 = 3400 \, \text{Hz}, \) and \( f_2 = 14500 \, \text{Hz} \).

Note that the screen splitting approach introduces significant signal degradation as the number of strips \( n \) increases, primarily due to reduced pixel intensity per strip and heightened inter-strip interference. For instance, while a single strip achieves an SNR of over 22~dB at 0.5 meters, splitting the screen into \( n = 4 \) strips could reduce the SNR to below 10~dB at the same distance, following the proportional signal attenuation observed. In a realistic setup, where the receiver is positioned close to the transmitter (e.g., within 1 meter), experimental evaluations indicate that the maximum usable number of strips \( n \) ranges between 1 and 4. Beyond this threshold, the SNR approaches the noise floor, leading to unreliable signal decoding and diminished data transmission efficiency. This trade-off emphasizes the balance required between increasing transmission capacity through OFDM and maintaining sufficient signal quality for effective reception.

\section{Conclusion}
In this work, we presented a method for generating sound using liquid crystal displays (LCDs) by leveraging the acoustic noise resulting from rapid pixel state transitions and the accompanying power fluctuations within the display circuitry. By carefully modulating these transitions, we demonstrated how standard LCD screens can be repurposed as sound-emitting devices without the need for conventional audio hardware. Our approach involved detailed algorithmic designs for signal generation and transmission, providing a structured framework for producing controllable audio frequencies. We conducted evaluations on various LCD screens, analyzing how factors such as resolution, receiver positioning, and distance influence signal quality. Our findings suggest that reliable sound generation is achievable at close range, with noticeable signal attenuation beyond a maximum effective distance of approximately 4.5 meters. The achievable data transmission speed was shown to be closely linked to the modulation technique and bit duration. Specifically, we outlined how shorter bit durations and optimized pixel clock settings contribute to increased transmission rates. Additionally, we demonstrated vertical screen splitting combined with Frequency-Shift Keying (FSK) modulation, which facilitated the concurrent transmission of multiple frequencies, thereby achieving an aggregate bitrate proportional to the number of screen regions. We also examined advanced modulation schemes, including Orthogonal Frequency-Division Multiplexing (OFDM), to enhance spectral efficiency and improve data throughput. By presenting detailed methodologies and experimental insights, we provide a reference point for further exploration of LCD-based acoustic communication systems. While the approach has limitations such as range, speed, and sensitivity to environmental factors, it may serve as a practical technique for audio generation in specific scenarios where conventional audio hardware is unavailable or intentionally disabled.

\bibliographystyle{ieeetran}
\bibliography{DISPLAY}

\end{document}